\documentclass[twocolumn,aps,prl,showpacs,floatfix,superscriptaddress]{revtex4}
\usepackage{times}
\usepackage{amsmath}
\usepackage{amssymb}
\usepackage{hyperref}
\usepackage{latexsym}
\usepackage{epsfig}


\begin{document}

\title{Exact Solution of Strongly Interacting Quasi-One-Dimensional Spinor Bose Gases}

\author{F. Deuretzbacher}
\email{fdeuretz@physnet.uni-hamburg.de}
\affiliation{I. Institut f\"ur Theoretische Physik, Universit\"at Hamburg, Jungiusstr. 9, 20355 Hamburg, Germany}

\author{K. Fredenhagen}
\email{klaus.fredenhagen@desy.de}
\affiliation{II. Institut f\"ur Theoretische Physik, Universit\"at Hamburg, Luruper Chaussee 149, 22761 Hamburg, Germany}

\author{D. Becker}
\affiliation{I. Institut f\"ur Theoretische Physik, Universit\"at Hamburg, Jungiusstr. 9, 20355 Hamburg, Germany}

\author{K. Bongs}
\affiliation{Institut f\"ur Laserphysik, Universit\"at Hamburg, Luruper Chaussee 149, 22761 Hamburg, Germany}
\affiliation{MUARC, School of Physics and Astronomy, University of Birmingham, Edgbaston, Birmingham B15 2TT, UK}

\author{K. Sengstock}
\affiliation{Institut f\"ur Laserphysik, Universit\"at Hamburg, Luruper Chaussee 149, 22761 Hamburg, Germany}

\author{D. Pfannkuche}
\affiliation{I. Institut f\"ur Theoretische Physik, Universit\"at Hamburg, Jungiusstr. 9, 20355 Hamburg, Germany}

\begin{abstract}
We present an exact analytical solution of the fundamental system of quasi-one-dimensional spin-1 bosons with infinite $\delta$-repulsion. The eigenfunctions are constructed from the wave functions of non-interacting spinless fermions, based on Girardeau's Fermi-Bose mapping, and from the wave functions of distinguishable spins. We show that the spinor bosons behave like a compound of non-interacting spinless fermions and non-interacting distinguishable spins. This duality is especially reflected in the spin densities and the energy spectrum. We find that the momentum distribution of the eigenstates depends on the symmetry of the spin function. Furthermore, we discuss the splitting of the ground state multiplet in the regime of large but finite repulsion.
\end{abstract}


\maketitle

{\it Introduction.---} Strong correlations are the basis of some of the most fascinating and important phenomena in many-body quantum systems. One intriguing only recently experimentally realized example~\cite{Paredes04, Kinoshita04} is the Tonks-Girardeau gas~\cite{Girardeau60}, in which a one-dimensional (1D) system of strongly interacting bosons acquires fermionic properties. Already in 1960 Girardeau found an elegant solution of these systems which maps the wave functions of non-interacting spinless fermions to that of spinless hard-core bosons. However, a thorough description of quasi-1D bosons needs to include the spin degrees of freedom. Surprisingly, thus far no Fermi-Bose mapping for these important systems exists. In current investigations Girardeau's Fermi-Bose mapping has been extended to fermionic Tonks-Girardeau gases~\cite{Cheon98, Girardeau04, Granger04, Bender05} and very recently also to mixtures~\cite{Girardeau07-1} and two-level atoms~\cite{2-Level_Atoms}. Ferromagnetic behavior in a strongly interacting two-component Bose gas has been found in the thermodynamic limit~\cite{Xi-Wen_Guan07}.

In this Letter we present for the first time an analytically exact solution of quasi-1D bosons with infinite $\delta$-repulsion, which includes the spin degrees of freedom. This opens up new possibilities for a precise manipulation and description of these fundamental systems. We show that the strongly interacting spinor bosons behave like a compound of non-interacting spinless fermions and non-interacting distinguishable spins which is especially reflected in the spin densities and the energy spectrum. These properties are very different from those of weakly interacting spinor Bose gases~\cite{Spinor_BECs}. Another key result is that in contrast to single-component systems the momentum distribution of the spinor bosons depends on the symmetry of the spin function.

{\it Model.---} The many-body Hamiltonian describing $N$ quasi-1D spin-1 bosons at zero magnetic field is given by the ${3^N \!\! \times \! 3^N}$ dimensional matrix~\cite{Spinor_BECs}
\begin{eqnarray} \label{H-model}
H & = & \sum_{i=1}^N \left[ -\frac{\hbar^2}{2m} \frac{\partial^2}{\partial z_i^2} + \frac{1}{2} m \omega^2 z_i^2 \right] \openone_\mathrm{spin} \nonumber \\
& & + \sum_{i<j} \delta(z_i - z_j) \left[ U_0 \openone_\mathrm{spin} + U_2 \vec f_i \cdot \vec f_j \right] ,
\end{eqnarray}
where $m$ is the mass of the bosons, $\omega$ is the axial trap frequency, $\openone_\mathrm{spin}$ is the identity matrix in spin space, $U_0$ and $U_2$ are the coupling constants of the spin-independent and spin-dependent interactions, and $\vec f_i$ denote the dimensionless spin-1 matrices of boson $i$. We consider a highly elongated trap where the transverse trap frequency $\omega_\bot$ is much larger than the axial trap frequency ${(\omega_\bot \gg \omega)}$ so that the transverse motion is restricted to zero-point oscillations. In this limit the effective coupling constants are given by ${U_i = 2 \hbar \omega_\bot c_i \gamma}$ (${i = 0,2}$) with ${\gamma = 1/(1-1.46 a_0 / \sqrt{2} l_\bot)}$~\cite{Olshanii98}, ${c_0 = (a_0 + 2a_2) / 3}$, and ${c_2 = (a_2 - a_0) / 3}$~\cite{Spinor_BECs}, where $a_0$ and $a_2$ are the s-wave scattering lengths and $l_\bot = \sqrt{\hbar / (m \omega_\bot)}$. We neglect p-wave scattering~\cite{Granger04}. Thus, $U_0$ and $U_2$ are proportional to the transverse trap frequency $\omega_\bot$ which has been used in experiments~\cite{Kinoshita04} to reach the strong interaction limit. The Hamiltonian (\ref{H-model}) conserves parity $\Pi$, total magnetization $F_z$, and total spin $\vec F^2$ ${\bigl( \vec F = \sum_i \vec f_i \bigr)}$. In the limit of infinite repulsion ${(U_0 = \infty)}$ we use a Fermi-Bose map to construct the eigenfunctions of (\ref{H-model}), whereas for large but finite repulsion we numerically diagonalize (\ref{H-model}) via a finite basis set approach~\cite{Deuretzbacher07}.

{\it Exact eigenstates in the limit of infinite repulsion.---} We first construct the exact eigenfunctions of the Hamiltonian (\ref{H-model}) in the limit of infinite repulsion ${(U_0 = \infty)}$. In this limit one can neglect the spin-dependent interaction and we set ${U_2 = 0}$. For simplicity we restrict our discussion to the ground states of spin-1 bosons. Because of the infinite $\delta$-repulsion the configuration space $\mathbb{R}^N$ decomposes into $N!$ sectors ${C_\pi = \{ (z_1, \ldots, z_N) \in \mathbb{R}^N , \ z_{\pi(1)} < \cdots < z_{\pi(N)} \}}$, where $\pi$ is an arbitrary permutation. At the boundaries of these sectors the wave function has to be zero ${[ \psi(z_1, \ldots, z_N) = 0}$ if ${z_i = z_j ]}$ and within these sectors the wave function has to obey the Schr\"odinger equation of non-interacting particles.

Let us disregard the spin of the wave function in a first step. The Fermi ground state $\psi_0^F$ of the problem above is given by the ground state Slater determinant of non-interacting spinless fermions. The ground state of spinless bosons $\psi_0^B$ is given by the absolute value of the Fermi ground state~\cite{Girardeau60}: ${\psi_0^B = \left| \psi_0^F \right|}$. There is no symmetry restriction to the wave function of distinguishable particles. Thus, any restriction of the Tonks-Girardeau ground state $\psi_0^B$ to an arbitrary sector $C_\pi$ is a non-symmetric ground state of $N$ spinless distinguishable particles which we denote by ${| \pi \rangle}$. Restrictions to different sectors are orthogonal. Thus, the ground state of $N$ spinless distinguishable particles is $N!$ times degenerate. An orthonormal basis of the space of ground states is therefore given by
\begin{equation} \label{basis-wf}
\langle z_1, \ldots ,z_N | \pi \rangle \! = \! \left\{
\begin{aligned}
& \mspace{-5mu} \sqrt{N!} \left| \psi^F_0 \right| \; \text{if} \; z_{\pi(1)} < \cdots < z_{\pi(N)} \\
& \mspace{-5mu} 0 \;
\text{otherwise.}
\end{aligned}
\right.
\end{equation}

Now we take the spin into account. The product of an arbitrary orbital function ${| \pi \rangle}$ with an arbitrary $N$-particle spin function ${| \chi \rangle}$ is an eigenfunction of Hamiltonian (\ref{H-model}), since $H$ is diagonal in spin space (we have neglected $U_2$). Thus, the wave function ${| \pi \rangle | \chi \rangle}$ describes distinguishable hard-core particles with spin. In order to describe bosons one has to symmetrize this non-symmetric wave function. One can directly construct all bosonic solutions from the spin functions by means of the unitary map
\begin{equation} \label{map}
W | \chi \rangle = \sqrt{N!} P_S \left( | \text{id} \rangle | \chi \rangle \right) ,
\end{equation}
where ${P_S = \frac{1}{N!} \sum_{\pi \epsilon S_N} U(\pi)}$ denotes the symmetric projection operator and id is the identical permutation. The action of the permutation operators ${U(\pi)}$ onto the spinor wave functions is given by
\begin{equation*}
U(\pi) | \pi^\prime \rangle | f_1, \ldots , f_N \rangle = | \pi \pi^\prime \rangle | f_{\pi^{-1}(1)}, \ldots , f_{\pi^{-1}(N)} \rangle ,
\end{equation*}
where ${\pi \pi^\prime}$ is the composition of the permutations $\pi$ and $\pi^\prime$, and ${f_i = \pm 1, 0}$ are the spin indices of the $i$th particle. Equation~(\ref{map}), as the main result of our Letter, provides an analytical and general extension of Girardeau's Fermi-Bose mapping to bosons with spin degrees of freedom. The map $W$ is isometric, and therefore, an orthonormal set of spin functions is directly mapped onto an orthonormal set of bosonic ground states. It is also bijective, and thus the degeneracy of the ground state equals the dimension of the $N$-particle spin space. Further, we can directly construct bosonic $F_z$ and $\vec F^2$ eigenfunctions ${W | \chi \rangle}$, since $F_z$ and $\vec F^2$ commute with $P_S$.

The above construction scheme can be extended to the excited states and to bosons with arbitrary spin. To obtain the excited states one has to replace ${\left| \psi^F_0 \right|}$ in Eq.~(\ref{basis-wf}) by ${A \psi^F_i}$, where $\psi^F_i$ is the $i$th eigenfunction of the non-interacting fermionic problem and $A$ is the ``unit antisymmetric function''~\cite{Girardeau60}. It follows immediately that the energy eigenvalues of the spin-1 bosons at zero magnetic field agree with the energies of spinless non-interacting fermions; i.e., the ground state energy is given by ${E_g = N^2/2 \, \hbar \omega}$ and the level spacing is ${\Delta E = 1 \, \hbar \omega}$. However, the degeneracy of each energy level is $3^N$ times larger than in the spinless case. A homogeneous magnetic field lifts the degeneracy of the levels and shifts the energy of each $F_z$ eigenstate ${W | f_1, \ldots, f_N \rangle}$ according to the Zeeman energy of the spin function ${| f_1, \ldots, f_N \rangle}$.

\begin{figure}

  \centering
  \includegraphics[width = 0.94\columnwidth]{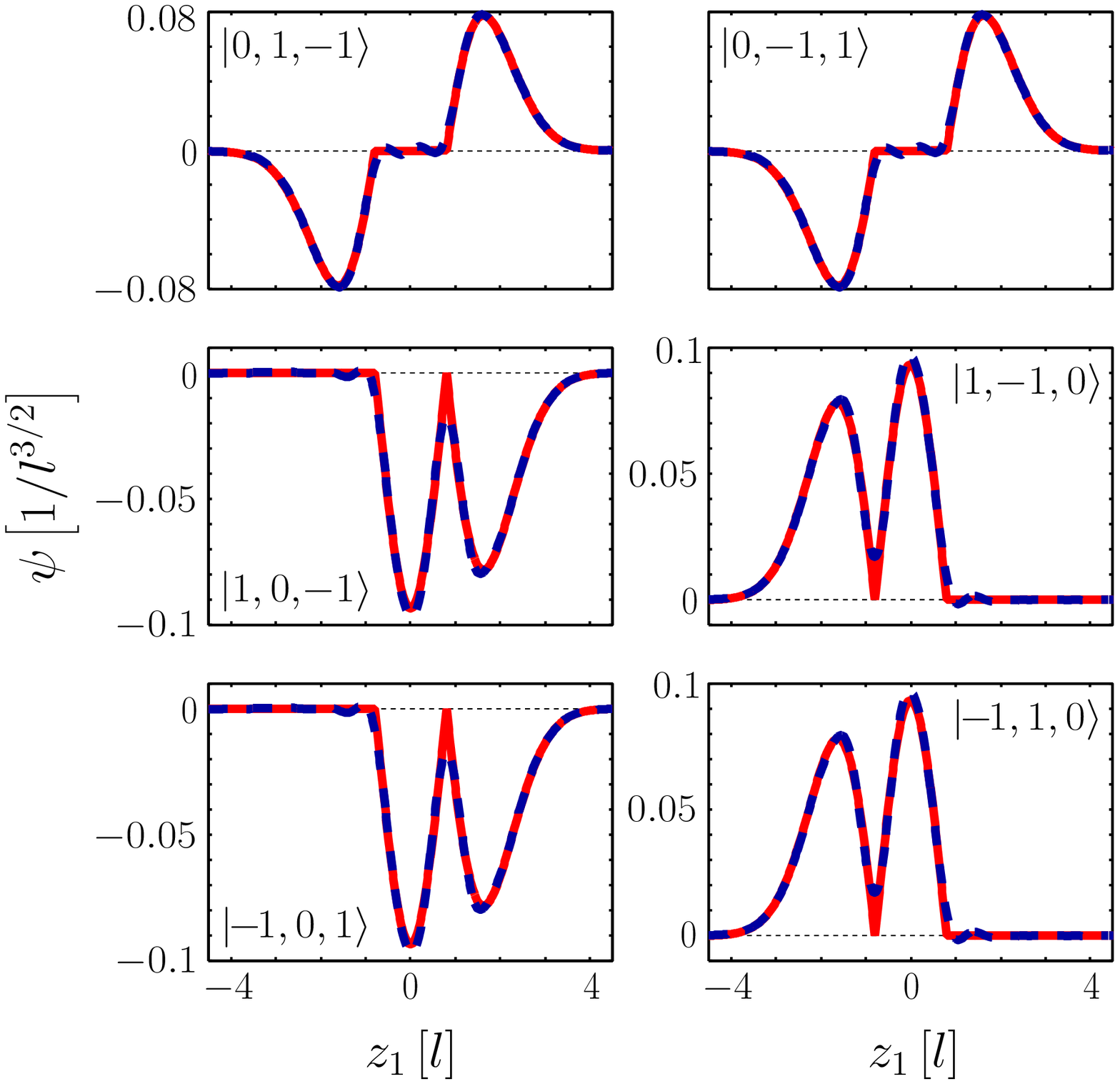}

  \caption{(color online) Cut through the non-zero spin-components of a bosonic 3-particle state (see text). The second and the third coordinate are fixed at ${z_2 = -0.8 \, l}$ and ${z_3 = 0.8 \, l}$. Shown is the exact wave function in the limit of infinite $\delta$-repulsion (red solid line) and the solution of a numerical diagonalizion of (\ref{H-model}) in the limit of large but finite repulsion (blue dashed line). \label{fig-wf}}
\end{figure}

{\it Large finite repulsion.---} In the following we analyze the structure of the ground state multiplet in the regime of large but finite repulsion. The results are based on a numerical diagonalization of (\ref{H-model}). In the limit of infinite repulsion, ${U_0 = \infty}$, the ground state is $3^N$ times degenerate. For realistic systems the spin-independent interaction $U_0$ is large and finite and the spin-dependent interaction $U_2$ is much smaller so that the ground state energy level is quasidegenerate. Let us first consider the case that $U_2$ is zero. Then, states with completely symmetric spin functions have lowest energy~\cite{Eisenberg_Lieb_02} whereas states with most antisymmetric spin functions have highest energy. The energy gap ${\Delta E_0}$, which arises from the symmetry of the spin function (which is closely related to the symmetry of the corresponding orbital functions), is proportional to ${\hbar \omega /\widetilde{U}_0}$ ${\bigl[ \, \widetilde{U}_i = U_i / (\hbar \omega l)}$ are dimensionless interaction strengths and ${l = \sqrt{\hbar / (m \omega} \; \bigr]}$. A similar energy structure has been discussed in Ref.~\cite{Girardeau07-1} for a two-component system. Let us now switch on $U_2$. Then, all states become $\vec F^2$ eigenstates and the states with completely symmetric spin functions have total spin ${F = N}$, ${N-2}$, ${N-4}$, ... \cite{Eisenberg_Lieb_02}. For ferromagnetic coupling, ${U_2 < 0}$, states with ${F = N}$ have lowest energy whereas for antiferromagnetic coupling, ${U_2 > 0}$, states with ${F = 1}$ (if $N$ is odd) or ${F = 0}$ (if $N$ is even) become ground states. The energy gap ${\Delta E_2}$, which arises from the spin-dependent interaction, is proportional to ${\hbar \omega \, \widetilde{U}_2 / \widetilde{U}_0^2}$, since the local correlation function $g_2$ is proportional to ${1/\widetilde{U}_0^2}$ in the limit of strong repulsion~\cite{Gangardt03} and ${\Delta E_2 = U_2 g_2}$.

For a sufficiently large repulsion, the quasidegenerate ground states are well approximated by the states (\ref{map}). As an example, Fig.~\ref{fig-wf} shows the 3-boson state $| \psi \rangle = W | \chi \rangle$ with $| \chi \rangle = 1/2 ( | 0, 1, -1 \rangle + | 0, -1, 1 \rangle - | 1, -1, 0 \rangle - | -1, 1, 0 \rangle )$ (red solid line), which is a $F_z$, $\vec F^2$, and parity eigenstate (${F_z = 0}$, ${F = 1}$, ${\Pi = -1}$), compared to the corresponding solution of a numerical diagonalization of (\ref{H-model}) with parameters ${U_0/l = 20 \, \hbar \omega}$ and ${U_2 = -U_0 / 2000}$ (blue dashed line). Both solutions are in excellent agreement, and thus the limit of infinite repulsion is practically reached at ${U_0/l = 20 \, \hbar \omega}$~\cite{Deuretzbacher07, Fermionization}. Note that at finite $U_0$ the state shown in Fig.~\ref{fig-wf} is an excited state within the ground state multiplet, since ${| \chi \rangle}$ is not completely symmetric and ${F < 3}$.

{\it Spin densities of the ground states.---} Apart from the energy spectrum, the dual nature of the hard-core bosons with spin can best be seen in the densities. While the total density is independent of the spin degrees of freedom and equal to that of non-interacting fermions, the spin density
\begin{eqnarray*}
\rho_{f_z}(z) & = & \sum_i \int dz_1 \ldots dz_N \sum_{f_1, \ldots, f_N} \delta(z - z_i) \delta_{f_z f_i} \\
& & \times |\psi_{f_1, \ldots, f_N}(z_1, \ldots, z_N)|^2
\end{eqnarray*}
resembles the one of a chain of localized, distinguishable spins. In the limit of infinite $\delta$-repulsion the wave function is given by ${| \psi \rangle = W | \chi \rangle}$ and one finds
\begin{equation} \label{spin-density}
\rho_{f_z}(z) = \sum_i p_{i}(f_z)\rho^{(i)}(z)
\end{equation}
with the probability $p_i(f_z)$ that the $i$th spin is equal to $f_z$,
\begin{equation*}
p_i(f_z) = \sum_{f_1, \ldots, f_N}| \langle f_1, \ldots, f_N | \chi \rangle |^2 \delta_{f_z f_i} ,
\end{equation*}
and the probability density $\rho^{(i)}(z)$ to find the $i$th particle of the system, restricted to the standard sector $C_\text{id}$, at point $z$,
\begin{eqnarray*}
\rho^{(i)}(z) & = & N! \int_{z_1 < \cdots < z_N} dz_1 \cdots dz_N |\psi^F_0(z_1, \ldots, z_N)|^2 \\
& & \times \delta(z-z_i) .
\end{eqnarray*}
An explicit calculation yields the following formula
\begin{eqnarray*}
\rho^{(i)}(z) & = & \frac{\partial}{\partial z} \Biggl( \sum_{k=0}^{N-i} \frac{(-1)^{N-i}(N-k-1)!}{(i-1)!(N-k-i)!k!} \\
& & \times \frac{\partial^k}{\partial\lambda^k}\mathrm{det}[B(z)-\lambda 1]|_{\lambda=0} \Biggr) ,
\end{eqnarray*}
where the ${N \times N}$-matrix $B(z)$ has entries $\beta_{ij}(z) = \int_{-\infty}^z dx \phi_i(x) \phi_j(x)$ with the single-particle eigenfunctions of the spinless problem $\phi_i$.

\begin{figure}

  \centering
  \includegraphics[width = \columnwidth]{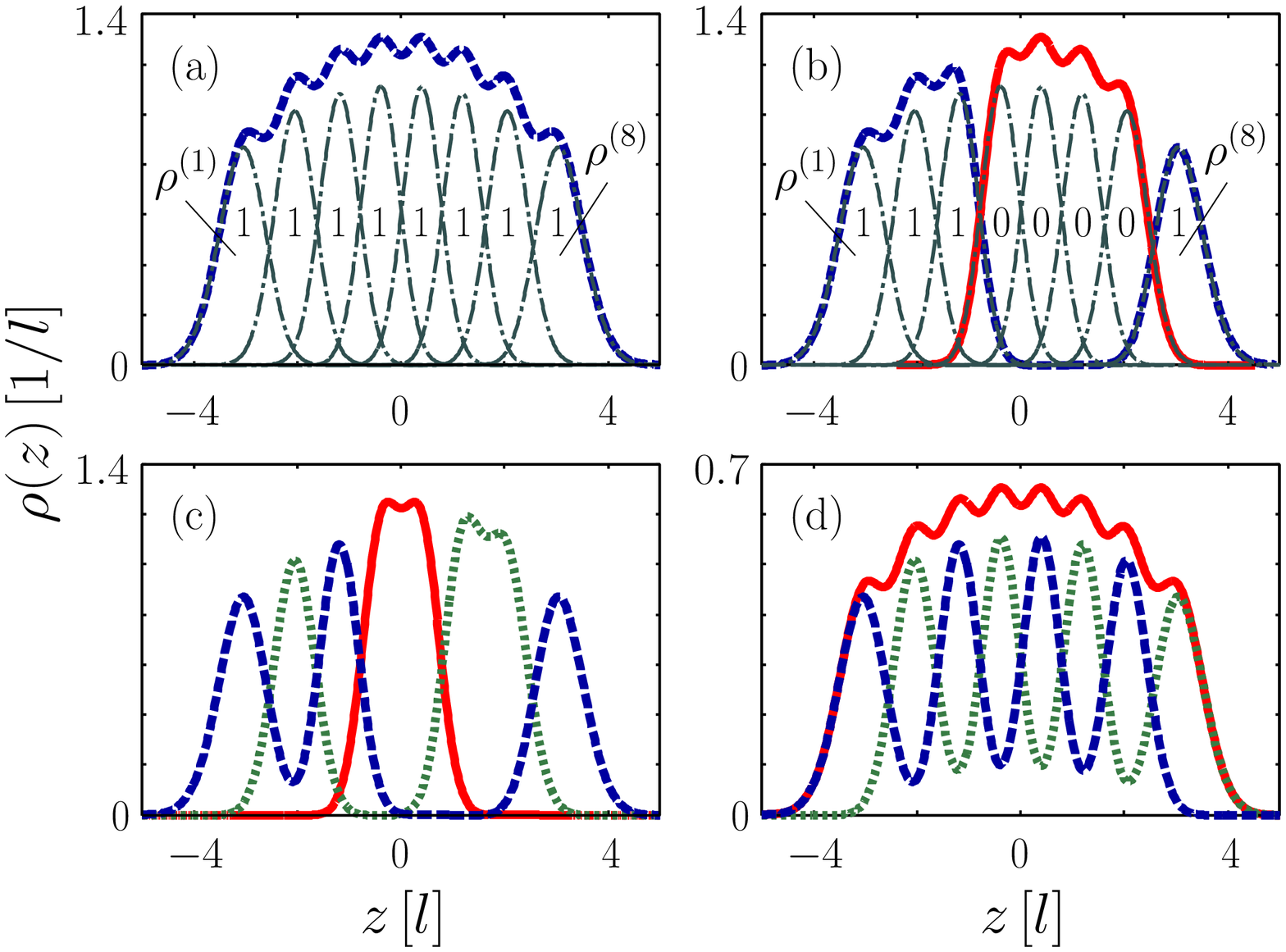}

  \caption{(color online) Spin densities of 8 spin-1 bosons in different ground states (see text). Shown are the densities $\rho^{(i)}$ (gray dash-dotted line, see text), and the components $\rho_0$ (red solid line), $\rho_1$ (blue dashed line) and $\rho_{-1}$ (green dotted line) of the spin density. The spin densities resemble the one of a chain of localized spins. \label{fig-densities}}
\end{figure}

Spin densities of selected ground states of 8 spin-1 bosons are shown in Fig.~\ref{fig-densities}. Since the total spin of each atom is one, the spin density has 3 components which correspond to ${f_z = \pm 1, 0}$, and which are drawn as a blue dashed ($\rho_1$), red solid ($\rho_0$), and a green dotted line ($\rho_{-1}$). By interpreting the square root of the densities $\rho^{(i)}$ (gray dash-dotted line in Figs.~\ref{fig-densities}a and b)  as the wave packet of an imaginary particle, ${\psi^{(i)} := \sqrt{\rho^{(i)}}}$, we obtain the intuitive picture of distinguishable localized particles which are arranged in a row along the $z$-axis, one after the other~\cite{Kinoshita04, Localization}. Each particle has a certain spin orientation $f_i$ according to the spin state ${| \chi \rangle = | f_1, f_2, \ldots \rangle}$. The spin density of the bosonic system now corresponds to the spin density of this row of distinguishable particles with the spin orientations given by ${| \chi \rangle}$.

Fig.~\ref{fig-densities}a shows the spin density of the spin-polarized state $W | \chi_+ \rangle = W | 1, 1, \ldots \rangle$. In the spin-polarized case all the probabilities $p_i(1) = 1$ and Eq.~(\ref{spin-density}) reduces to: $\rho_1 = \sum_i \rho^{(i)} = \rho_\mathrm{Fermi}$. Fig.~\ref{fig-densities}b shows the spin density of the ground state ${W | 1, 1, 1, 0, 0, 0, 0, 1 \rangle}$. Similarly one has to add the particle densities $\rho^{(1)}-\rho^{(3)}$ and $\rho^{(8)}$ to the component $\rho_1$, and the particle densities $\rho^{(4)}-\rho^{(7)}$ to the component $\rho_0$ of the spin density. Finally, Figs.~\ref{fig-densities}c and d show the spin densities of the ground state ${W | 1, -1, 1, 0, 0, -1, -1, 1 \rangle}$ and the superposition state ${W \frac{1}{\sqrt{2}} \left( | 0, 0, \ldots \rangle + | 1, -1, 1, -1, \ldots \rangle \right)}$ respectively.

\begin{figure}

  \centering
  \includegraphics[width = \columnwidth]{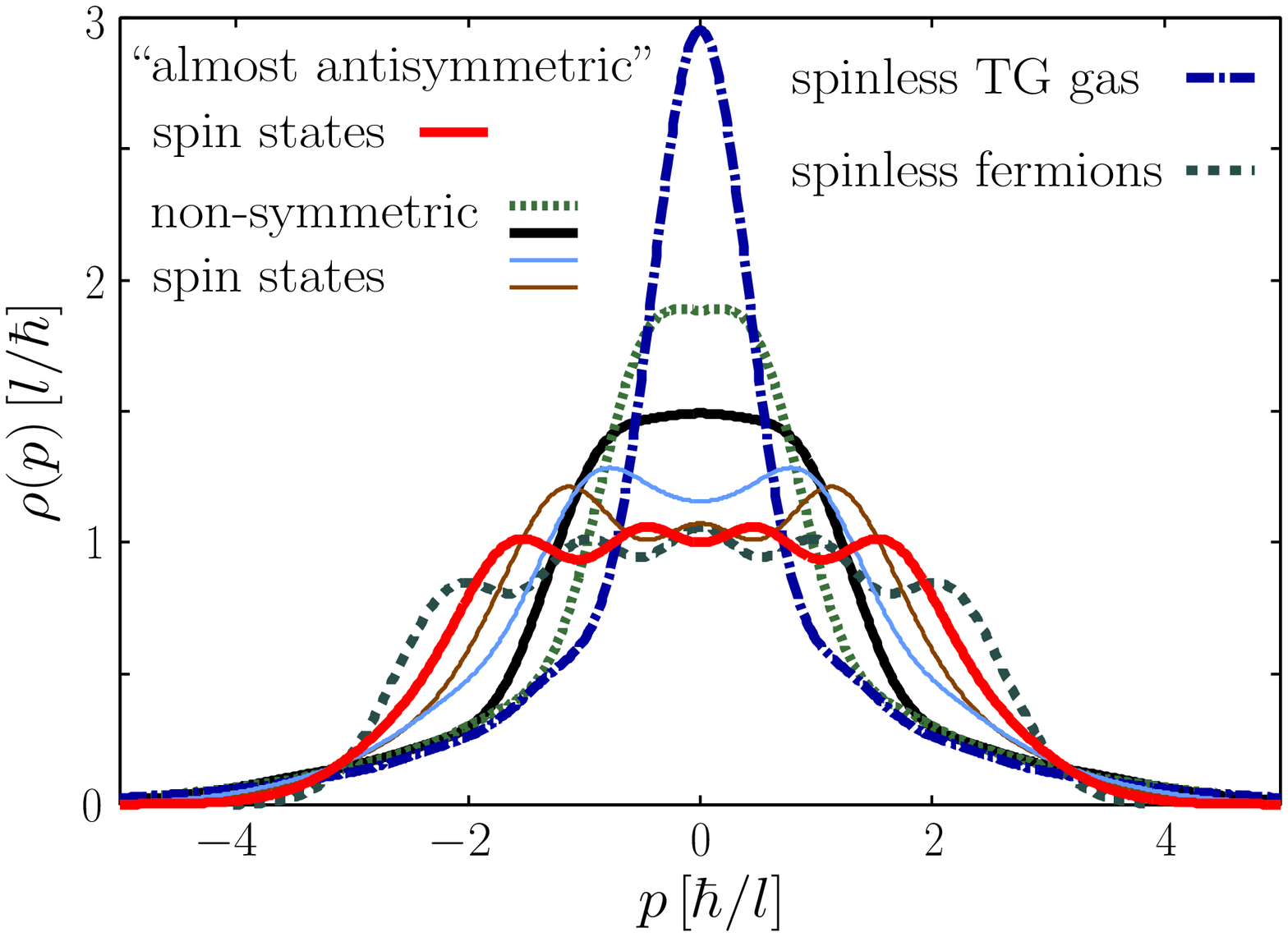}

  \caption{(color online) Momentum distributions of 5 spin-1 bosons in different ground states. The gray dashed line shows for comparison the momentum distribution of 5 non-interacting fermions. The shape of the momentum distribution depends on the symmetry of the spin function. \label{fig-momentum-distributions}}
\end{figure}

{\it Momentum distributions of the ground states.---} One of the most important experimentally accessible quantities is the momentum distribution of the spinor bosons, given by
\begin{eqnarray*}
\rho(p) & = & \sum_i \int dp_1 \ldots dp_N \sum_{f_1, \ldots, f_N} \delta(p - p_i) \\
& & \times |\psi_{f_1, \ldots, f_N}(p_1, \ldots, p_N)|^2 .
\end{eqnarray*}
Fig.~\ref{fig-momentum-distributions} shows selected momentum distributions of 5 spinor bosons in their degenerate ground states, obtained from a numerical diagonalization of (\ref{H-model}). For comparison we have also plotted the momentum distribution of 5 non-interacting fermions (gray dashed line). It turns out that the shape of the momentum distribution depends on the symmetry of the spin function so that different ground states can have completely different momentum distributions. States with a completely symmetric spin function have a completely symmetric orbital function, ${W | \chi_s \rangle = \left| \psi^F_0 \right| | \chi_s \rangle}$, which is given by the usual Tonks-Girardeau wave function. The momentum distribution of these states is equal to the one of a spinless Tonks-Girardeau gas (blue dash-dotted line) which exhibits a pronounced zero-momentum peak and long-range, high-momentum tails~\cite{Deuretzbacher07, Momentum_Distribution}. The other extreme case is given by a flat and broad momentum distribution which resembles the fermionic one (red solid line). In the case of 2 and 3 spin-1 bosons some spin functions ${| \chi_a \rangle}$ can be completely antisymmetric and thus the corresponding ground state is given by ${W | \chi_a \rangle = \psi_0^F | \chi_a \rangle}$ so that its momentum distribution is equal to the one of spinless fermions. One cannot construct completely antisymmetric spin functions with more than 3 spin-1 particles. However, it is possible to construct non-symmetric spin functions which are ``almost antisymmetric'' (see Young's Tableaux~\cite{FiniteGroups}) resulting in momentum distributions which are almost fermionic. We believe, due to group theoretical arguments~\cite{FiniteGroups}, that this broadening and flattening of some momentum distributions saturates for large $N$.

{\it Preparation and manipulation of the states.---} Spin polarized Tonks-Girardeau gases have been prepared with ultracold $^{87}$Rb atoms in optical lattices~\cite{Paredes04, Kinoshita04}. To prepare states with non-symmetric spin functions we suggest to apply a radio frequency together with a magnetic field gradient so that the spin of each atom is rotated into a different final state depending on its position.

{\it Summary.---} We have constructed the exact eigenstates of the fundamental system of quasi-1D spin-1 bosons with infinite $\delta$-repulsion by means of a Fermi-Bose map. The construction scheme and the formula for the spin densities are valid independent of the particle number, the spin of the bosons, and the (spin-independent) trapping potential. From the exact solutions we have determined the energy spectrum and the spin densities of the ground states. The momentum distribution of the eigenstates turned out to be dependent on the symmetry of the spin functions. Some ground states have a momentum distribution which is much broader and flatter than for a spinless Tonks-Girardeau gas. For large but finite repulsion we have discussed the level structure of the ground state multiplet. Due to the similarity with distinguishable spins the spinor Tonks-Girardeau gas might be a promising candidate for novel quantum computation schemes.

{\it Acknowledgments.---} We thank K.~Plassmeier for his help on various numerical problems and we acknowledge fruitful discussions with C.~Becker, X.-W.~Guan, J.~Kronj\"ager, D.~S.~L\"uhmann, M.~Oberthaler, S.~Reimann, A.~I.~Streltsov, and S.~Z\"ollner. K.~B. thanks EPSRC for financial support in Grant No. EP/E036473/1.


\begin{thebibliography}{22}

\bibitem{Paredes04} B.~Paredes {\it et al.}, Nature (London) {\bf 429}, 277 (2004).

\bibitem{Kinoshita04} T.~Kinoshita, T.~Wenger, and D.~S. Weiss, Science {\bf 305}, 1125 (2004).

\bibitem{Girardeau60} M.~D.~Girardeau, J. Math. Phys. {\bf 1}, 516 (1960); Phys. Rev. {\bf 139}, B500 (1965), Secs. 2, 3, and 6.

\bibitem{Cheon98} T.~Cheon and T.~Shigehara, Phys. Lett. A {\bf 243}, 111 (1998); Phys. Rev. Lett. {\bf 82}, 2536 (1999).

\bibitem{Girardeau04} M.~D.~Girardeau and M.~Olshanii, Phys. Rev. A {\bf 70}, 023608 (2004).

\bibitem{Granger04} B.~E.~Granger and D.~Blume, Phys. Rev. Lett. {\bf 92}, 133202 (2004)

\bibitem{Bender05} S.~A.~Bender, K.~D.~Erker, and B.~E.~Granger, Phys. Rev. Lett. {\bf 95}, 230404 (2005).

\bibitem{Girardeau07-1} M.~D.~Girardeau and A.~Minguzzi, Phys. Rev. Lett. {\bf 99}, 230402 (2007).

\bibitem{2-Level_Atoms} S.~V.~Mousavi, A.~del~Campo, I.~Lizuain, and J.~G.~Muga, Phys. Rev. A {\bf 76}, 033607 (2007); M.~D.~Girardeau, arXiv:0707.1884.

\bibitem{Xi-Wen_Guan07} X.-W.~Guan, M.~T.~Batchelor, and M.~Takahashi, Phys. Rev. A {\bf 76}, 043617 (2007).

\bibitem{Spinor_BECs} T.-L.~Ho, Phys. Rev. Lett. {\bf 81}, 742 (1998); T.~Ohmi and K.~Machida, J. Phys. Soc. Jpn. {\bf 67}, 1822 (1998); C.~K.~Law, H.~Pu, and N.~P.~Bigelow, Phys. Rev. Lett. {\bf 81}, 5257 (1998); J.~Stenger {\it et al.}, Nature (London) {\bf 396}, 345 (1998); M.~R.~Matthews {\it et al.}, Phys. Rev. Lett. {\bf 81}, 243, (1998).

\bibitem{Olshanii98} M.~Olshanii, Phys. Rev. Lett. {\bf 81}, 938 (1998).

\bibitem{Deuretzbacher07} F.~Deuretzbacher, K.~Bongs, K.~Sengstock, and D.~Pfannkuche, Phys. Rev. A {\bf 75}, 013614 (2007).

\bibitem{Eisenberg_Lieb_02} E.~Eisenberg and E.~H.~Lieb, Phys. Rev. Lett. {\bf 89}, 220403 (2002).

\bibitem{Gangardt03} D.~M.~Gangardt and G.~V.~Shlyapnikov, Phys. Rev. Lett. {\bf 90}, 010401 (2003).

\bibitem{Fermionization} Y. Hao, Y. Zhang, J.~Q. Liang, and S. Chen, Phys. Rev. A {\bf 73}, 063617 (2006); S.~Z\"ollner, H.-D.~Meyer, and P.~Schmelcher, Phys. Rev. A {\bf 74}, 053612 (2006).

\bibitem{Localization} I.~Romanovsky, C.~Yannouleas, and U.~Landman, Phys. Rev. Lett. {\bf 93}, 230405 (2004); O.~E.~Alon and L.~S.~Cederbaum, Phys. Rev. Lett. {\bf 95}, 140402 (2005).

\bibitem{Momentum_Distribution} M.~D.~Girardeau, E.~M.~Wright, and J.~M.~Triscari, Phys. Rev. A {\bf 63}, 033601 (2001); T.~Papenbrock, Phys. Rev. A {\bf 67}, 041601(R) (2003); S.~Z\"ollner, H.-D.~Meyer, and P.~Schmelcher, Phys. Rev. A {\bf 74}, 063611 (2006).

\bibitem{FiniteGroups} J.~J.~Sakurai, {\it Modern Quantum Mechanics} (Addison-Wesley, Reading, MA, 1994), Chap. 6.5.

\end{thebibliography}
\end{document}